\documentclass[aip,pof,preprint,floatfix]{revtex4-2} 

\usepackage[utf8]{inputenc}
\usepackage{graphicx}
\usepackage{amsmath,amssymb,mathtools}
\usepackage{bm}            
\usepackage{siunitx}       
\usepackage{dcolumn}       

\usepackage{multirow}
\usepackage{enumitem}
\usepackage{afterpage}

\usepackage{xcolor}

\usepackage{pifont}
\usepackage{soul}

\usepackage{nomencl}
\makenomenclature
\usepackage{etoolbox}
\renewcommand\nomgroup[1]{%
  \item[\bfseries
    \ifstrequal{#1}{A}{Acronyms}{%
    \ifstrequal{#1}{B}{Symbols}{}}%
  ]}

\usepackage{hyperref}
\usepackage{hypcap}

\graphicspath{{figures/}}

\begin{document}

\pagenumbering{roman}

\mbox{}

\nomenclature[B]{$\mu$}{dynamic viscosity}
\nomenclature[B]{$\theta$}{contact angle}
\nomenclature[B]{$\sigma$}{interfacial tension (IFT)}
\nomenclature[B]{$\sigma_{cut-oll}$}{cut-oll interfacial tension}
\nomenclature[B]{$r$}{pore throat radius}
\nomenclature[B]{$r_{mean}$}{mean pore throat radius}
\nomenclature[B]{$P$}{Capillary pressure}
\nomenclature[B]{$R^H$}{Hydraulic resistance}
\nomenclature[B]{$L$}{pore body length}
\nomenclature[B]{$Q$}{Flux}
\nomenclature[B]{$\triangle p$}{pressure drop}
\nomenclature[B]{$P_{in}$}{driving/inlet pressure}
\nomenclature[B]{$P_{out}$}{outlet pressure}
\nomenclature[B]{$V$}{velocity}
\nomenclature[B]{$\beta$}{tuning parameter}
\nomenclature[B]{$y,y0,A$}{fitting parameters}





\title{Spatio-temporal dynamics of surfactant driven secondary invasion in Gaussian pore networks}

\author{Debanik Bhattacharjee}
\author{Guy Z. Ramon}
\author{Yaniv Edery}
\email{Corresponding author: yanivedery@technion.ac.il}
 
\affiliation{Department of Civil \& Environmental Engineering Technion - Israel Institute of Technology Haifa 32000 Israel}%

\begin{abstract}
\pagenumbering{arabic}
Capillarity-dominated two-phase displacement in porous media often continues beyond the initial invasion-percolation (IP) breakthrough, as surfactants alter interfacial properties and reopen pathways once sealed by capillary forces. This study examines such secondary invasion, where adsorption-driven reductions in interfacial tension and contact-angle shifts lower entry thresholds in yet uninvaded throats, enabling further displacement at a fixed inlet pressure. To capture this process, we employ a time-dependent pore-network framework that couples IP with a reduced-order transport-adsorption module. Local fluxes are governed by Poiseuille flow, interfacial adsorption follows a Langmuir isotherm, and wettability evolution is modeled through a calibrated phenomenological relation. Heterogeneity is prescribed by Gaussian throat-size distributions whose variance controls structural disorder. The resulting invasion trajectories are sigmoidal, consistent with Gaussian cumulative statistics, indicating that surfactant mass-transfer kinetics and network variance primarily rescale invasion timescales while preserving the overall functional form. The framework thus connects interfacial conditioning to time-varying capillary thresholds and reveals how surfactant-mediated processes govern post-breakthrough dynamics in heterogeneous porous systems.

\end{abstract}

\maketitle
\section{Introduction}
Immiscible two-phase displacement in complex pore spaces underpins several technology areas. 
For example, in geological carbon storage, CO$_2$-brine displacement, residual trapping, and plume evolution are governed by capillary entry thresholds, wettability, and network connectivity \cite{Krevor2015,Chalbaud2009,Basirat2017,Aggelopoulos2011,Liu2024}.
Similar pore-scale physics arise in underground hydrogen storage, where drainage-imbibition cycling, contact-angle hysteresis, and capillary trapping modulate capacity and deliverability \cite{Zivar2021,Sambo2022,Muhammed2022,Okoroafor2022,Davoodi2025}. 
Beyond the subsurface, water management in proton-exchange-membrane fuel cells is controlled by capillary-driven invasion and percolation within gas-diffusion layers, coupling microstructure and wettability to liquid-water removal \cite{Sinha2007,Gostick2006,Zhou2010,Zhan2019,Wang2021,Falcao2009}.
Analogous displacement phenomena occur in liquid-composite molding of fibrous preforms, where interfacial forces and pore-network topology govern front stability and void formation during infiltration \cite{Breard2003,Park2011,DeValve2013,Pierce2017,Lu2024}. 
The aforementioned examples all share a mechanistic core -
 capillary entry conditions, connectivity, and wettability.

In capillary-dominated two-phase displacement, a pore-network model (PNM) simplifies pore space into nodes and throats, with local rules that encode geometry and wettability. 
PNMs capture interfacial entry conditions and phase configurations at the element level while enforcing global mass conservation, enabling predictive links between pore-scale mechanics and macroscopic invasion patterns \cite{Blunt2001}. 
Within this framework, invasion proceeds by admitting menisci in ascending order of their capillary entry thresholds, yielding a quasi-static, capillarity-controlled growth process that is well described by an invasion-percolation algorithm when viscous and buoyant forces are negligible. The practical utility of this approach has been
demonstrated across diverse media, including gas diffusion layers where invasion-percolation pathfinding coupled to pore-network flow accurately reproduces steady liquid distributions \cite{Lee2010}. 

Most IP/PNM formulations in the capillary limit assume fixed interfacial tension and contact angle and therefore cannot capture the continued, constant-pressure invasion observed when surfactants adsorb and reduce $\sigma\cos\alpha$.
We close this gap by coupling a standard pore-network model (PNM) to simplified transport-adsorption modules that update capillary entry thresholds over time.
In the quasi-static limit adopted here, a meniscus advances when its local entry pressure $p_i$ falls below the imposed inlet pressure $p_{\mathrm{in}}$; ranking the throats by $\{p_i\}$ yields a capillarity-controlled IP sequence that establishes the initial inlet-outlet path fixed at $p_{\mathrm{in}}$.
Once connected, we solve viscous flow only within the invaded cluster to obtain local interfacial fluxes, compute adsorption-driven updates of $\sigma$ and $\alpha$, and admit further invasion whenever the reduced thresholds are exceeded.
This produces a \emph{secondary} invasion driven by interfacial change rather than by any increase in the global pressure drop, while retaining the geometry-wettability links characteristic of PNMs \cite{Blunt2001}.

This formulation remains simple yet extensible: viscous-coupling corrections can be introduced when needed \cite{Xie2017}; dynamic rules can capture rate effects such as spontaneous imbibition and corner-flow participation \cite{Qin2019}; and heterogeneity can be represented either statistically or via image-based network extraction, including fractured materials \cite{Jiang2017}. 
We impose pore-scale heterogeneity through a stochastic Gaussian distribution of pore-throat radii with fixed mean and ensemble-specific variance. This simple network specification provides a one-parameter heterogeneity knob while keeping topology and statistics transparent. 
The forthcoming analysis quantifies how adsorption-driven changes in interfacial tension and wettability perturb an otherwise fixed IP pathway and induce additional, time-dependent advance of the invading phase in heterogeneous networks. 

Our principal finding is that the rate of secondary invasion is set by a mass-transfer timescale that shortens with stronger interfacial exchange (faster adsorption, higher diffusivity, greater local velocities) and lengthens as transport paths grow longer (i.e., at lower surface-area-to-volume). 
Increasing the variance of the pore-throat distribution raises the surface area to volume ratio, and exposes more reactive interfaces, thereby accelerating invasion even under a constant global pressure drop. 
Consequently, the invaded fraction follows a smooth, sigmoidal trajectory consistent with a Gaussian cumulative distribution. Heterogeneity and surfactant properties primarily stretch or compress the time axis, while leaving the functional form and the long-time limit unchanged. 
These observations place the system in a capillary-dominated, quasi-static regime: invasion steps are triggered by time-dependent reductions in entry thresholds at fixed pressure, whereas viscous pressure drops are used only to determine local velocities and thus the mass-transfer rate. In this regime, capillarity controls the mechanics of advance, and viscosity controls the kinetics of exchange.

\section{Methodology}

The pore-network, surfactant-transport, and adsorption models employed here follow the reduced-order formulation established in our previous work \cite{Bhattacharjee2025}. That earlier study focused on how non-Gaussian heterogeneity and surfactant properties influence the coupling between Laplace-pressure scaling and mass-transfer kinetics. In contrast, the present work extends the framework toward time-resolved secondary invasion, capturing the spatio-temporal evolution of the invaded fraction, quantifying how adsorption-driven interfacial changes modify capillary thresholds, and revealing sigmoidal scaling relations that characterize invasion trajectories. While the core model is retained for consistency and reproducibility, the emphasis here lies on new analyses, mechanisms, and physical insights that emerge from the temporal evolution of surfactant-mediated invasion.

\subsection{Pore Network Model (PNM)}

\begin{figure*}
\centering
\includegraphics[width=1.05\linewidth]{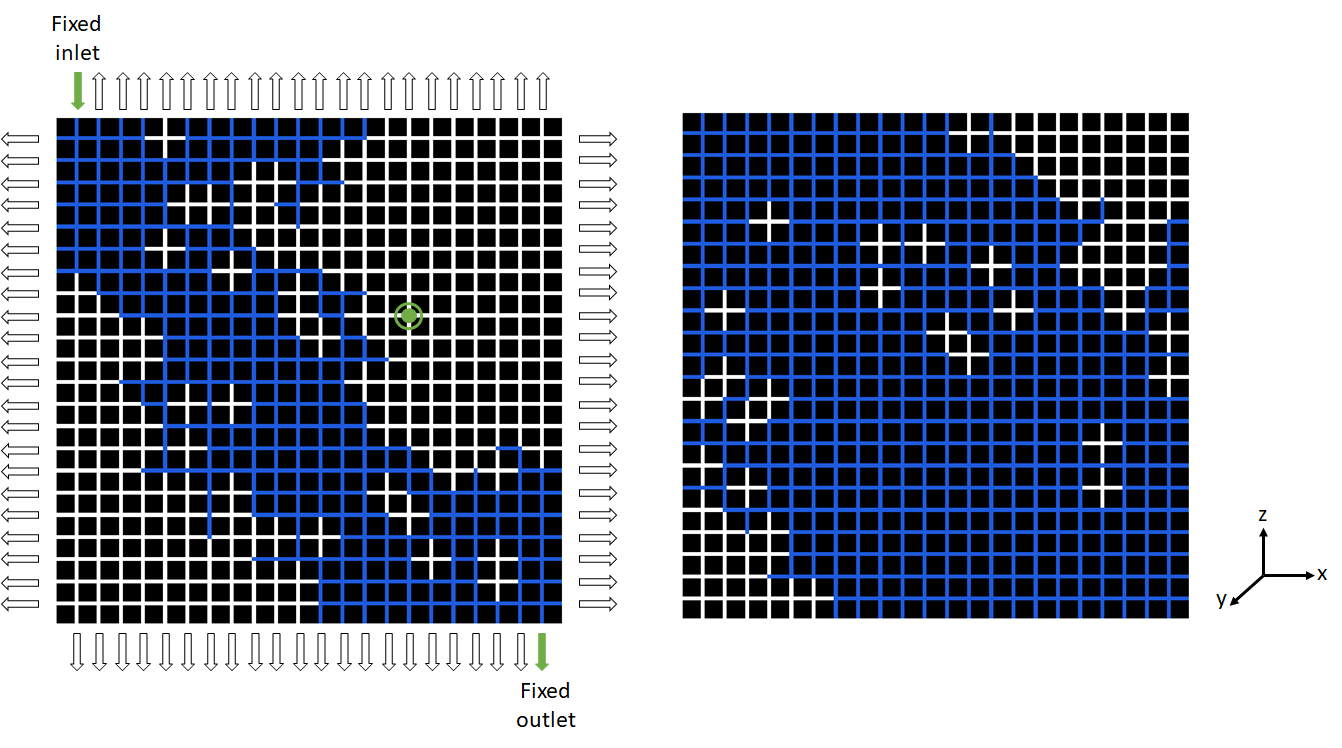}
\caption{Top-view of illustrative $22\times22$ micropillar networks showing baseline invasion-percolation (primary invasion) under constant inlet pressure, before surfactant effects. Throat radii are sampled from a Gaussian distribution with fixed mean and ensemble-specific variance. Open-boundary conditions are used; the inlet is fixed at the top-left and the outlet at the bottom-right (boundary arrows shown). Blue bonds delineate the first inlet-outlet connected path. The two panels are independent realizations with identical statistics, yielding distinct percolation paths that highlight the stochasticity of the pore-size heterogeneity. In the left panel, the green marker indicates a clustered region connected to the top boundary i.e., the clustered fluid can escape through the top. The trapped ganglia are apparent even as the interfacial tension is reduced, owing to the pore heterogeneity and local pressure drop. Axes indicate orientation; the view is along $+z$ (top view).}
\label{fig:1} 
\end{figure*}

\subsubsection{Geometry}

Pore networks are known to have multiple representations. 
Here, we adopt a two-dimensional pillar-lattice geometry (Fig.~\ref{fig:1}), in which black squares denote pillars. 
Intersections, where four pillars meet act as pore bodies, and the edges between neighboring pillars define pore throats, while their throat sizes are drawn from a Gaussian distribution, and the connectivity is specified by an adjacency matrix. The mean throat width is $50\,\mu\mathrm{m}$, with variance in the range $5$-$300\,\mu\mathrm{m}^2$.

We consider drainage, where a non-wetting water phase displaces a wetting oil in an hydrophobic network; fluid properties are summarized in Table~\ref{table:1}. For computational efficiency, water is injected at a fixed inlet node and collected at a fixed outlet node (highlighted with green arrows in Fig.~\ref{fig:1}), while the oil is permitted to drain through all other boundary nodes. 
These boundary conditions are applied across all realizations; complementary simulations with restricted outlets for the defending phase exhibited the same qualitative behavior. The inlet pressure is prescribed to be finite and nonzero, the outlet pressure is held at atmospheric, and invasion pathways are reconstructed directly from the adjacency matrix.

\subsubsection{Quasi-static invasion-percolation}
\label{subsubsection:qip-paper1}

Immiscible displacement is simulated using a quasi-static invasion-percolation (IP) scheme on a randomized two-dimensional pore network (Fig.~\ref{fig:1}). 
Each throat is modeled as a rectangular channel, so capillary entry and viscous losses follow standard rectangular-geometry expressions \cite{Juncker2002}. The capillary entry pressure at the throat $i$ is
\begin{equation}
\mathrm{{p}_i} \;=\; -\,\sigma\!\left(\frac{\cos\alpha_b+\cos\alpha_t}{\mathrm{h}}+\frac{\cos\alpha_l+\cos\alpha_r}{\mathrm{r_i}}\right),
\label{eq:Laplace}
\end{equation}
and the single-throat hydraulic resistance is
\begin{equation}
\mathrm{R}_{\mathrm{i}}^{\mathrm{H}} \;=\;
\frac{12\,\mathrm{\mu}\,\mathrm{L}}
{\mathrm{h}^{3}\,\mathrm{r}_{\mathrm{i}}\!\left(1 - 0.63\,\frac{\mathrm{h}}{\mathrm{r}_{\mathrm{i}}}\right)}
\end{equation}
\label{eq:Resistance}
where $\sigma$ is the interfacial tension, $\mu$ the dynamic viscosity, $\mathrm{L}$ the throat (channel) length, $\mathrm{h}$ the channel height, $\mathrm{r_i}$ the channel width, and $\alpha_b,\alpha_t,\alpha_l,\alpha_r$ the apparent contact angles on the four walls. (The prefactor $12$ reflects the parallel-plate limit for shallow rectangular microchannels and is consistent with rectangular-channel Poiseuille theory and experiments~\cite{Ichikawa2004,Park2008,Kim2016,Qi2008,Ergu2009}.)

To initialize the IP process, we sort the capillary thresholds $\{\mathrm{p_i}\}$ in ascending order and define the entry/inlet pressure as the smallest value for which the subset of throats with $\mathrm{p_i}$ smaller than this value forms a connected path from the fixed single inlet to the fixed single outlet. 
If no such value exists, the realization is excluded. This first connected configuration defines \emph{breakthrough}. Beyond breakthrough, the same (previously static) inlet pressure is interpreted as a hydraulic head that drives viscous flow through the formed conduit.

\subsubsection{Hydrodynamic flow}

After a connected path from inlet to outlet is identified, we compute the viscous pressure distribution and associated fluxes along the invaded network. 
Mass conservation is imposed at each node through a mass balance analogous to Kirchoff's law, which yields the pressure drops $\mathrm{\triangle P_{ij}}$ across connected nodes. Using the hydraulic resistances $\mathrm{R}_{\mathrm{i}\mathrm{j}}^{\mathrm{H}}$ defined in Eq.~\ref{eq:Resistance}, the edge fluxes follow a Hagen-Poiseuille relation, and the mean velocity is obtained by dividing the flux by the local cross-sectional area $\mathrm{hr_i}$. 
During this stage, only the invading phase carries flow; the displaced (oil) phase is taken as hydrodynamically inactive, consistent with the quasi-static IP setting.

\begin{equation}
\mathrm{\sum_{j}\frac{P_i}{R_{i}^H}}=0
\label{eq:Mass_Conservation}
\end{equation}
\begin{equation}
\mathrm{Q_{ij}} = \mathrm{\frac{\delta P_{ij}}{R_{ij}^{H}}}
\label{eq:Flux}
\end{equation}
\begin{equation}
\mathrm{V_{ij}} = \mathrm{\frac{Q_{ij}}{h r_i}}
\label{eq:Velocity}
\end{equation}

\begin{table*}
\centering
\includegraphics[width=1.0\linewidth]{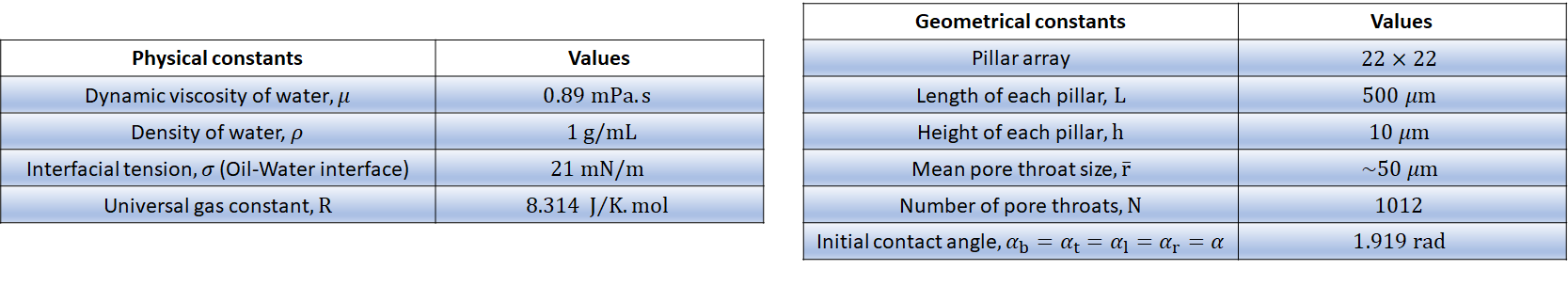}
\caption{Physical and geometrical constants used in all simulations. Unless otherwise noted, these values are held fixed across ensembles; variability arises from the stochastic pore-size distribution and (when applicable) surfactant parameters.}
\label{table:1} 
\end{table*}

\subsection{Surfactant Transport Model (STM)}
\label{subsection:STM}

Once the invasion pathway is established, surfactant is injected with the invading phase and transported toward available interfaces along the formed conduit. 
To represent near-interface transport in a manner consistent with advection-diffusion on the network, we adopt a first-order approximation in which the interfacial concentration relaxes toward the local bulk value:
\begin{equation}
\mathrm{\frac{dC_I}{dt}=\frac{k_I}{L_{I}}\,(C_B - C_I)},
\label{eq:MT1}
\end{equation}
where $\mathrm{C_B}$ is the throat-centered bulk concentration, $\mathrm{C_I}$ is the near-interface concentration, $\mathrm{k_I}$ is a mass-transfer coefficient, and $\mathrm{L_{I}=V/S}$ is the local volume-to-area ratio of the rectangular channel segment. For a throat of width encapsulating an interface, $\mathrm{r_I}$, height $\mathrm{h}$, and channel length $\mathrm{L}$, $\mathrm{V_{I}=r_I h\,L}$ and $\mathrm{S_{SF,I}=2(r_I{+}h)\,L}$, hence $\mathrm{L_{I} = r_I h/[2(r_I{+}h)]}$. 
Interfaces are considered at the throat centers (narrowest sections), consistent with the invasion geometry.

With the initial condition $\mathrm{C_I(t{=}0)=0}$, integration of Eq.~\ref{eq:MT1} yields
\begin{equation}
\mathrm{C_I = C_B\!\left(1-\exp\!\left[-\frac{k_I}{L_{I}}\,t\right]\right)}.
\label{eq:MT2}
\end{equation}

To estimate the local mass-transfer coefficient at the interface $\mathrm{I}$, we employ a mass transfer correlation of the form
\begin{equation}
\mathrm{Sh_I = Pe_I^{1/3} = \frac{k_I}{D/L_I}},
\label{eq:km}
\end{equation}
where $\mathrm{Sh_I}$ and $\mathrm{Pe_I}$ are the local Sherwood and Péclet numbers, and $D$ is the diffusivity.

Because both solid-fluid (SF) and fluid-fluid (FF) interfaces may be present, the near-interface concentration is represented as
\begin{equation}
\mathrm{C_{I}=n_{ISF}C_{ISF}+n_{IFF}C_{IFF},}
\quad
\mathrm{n_{ISF}+n_{IFF}=1.0},
\label{eq:Interface_Conc}
\end{equation}
where $\mathrm{n_{ISF}}$ and $\mathrm{n_{IFF}}$ are the interfacial fractions associated with SF and FF interfaces, respectively, and $\mathrm{C_{ISF}}$ and $\mathrm{C_{IFF}}$ are the corresponding near-interface concentrations.

\subsection{Surfactant Adsorption Model (SAM)}

The STM provides local bulk and near-interface concentrations along the invaded pathway. 
The Surfactant Adsorption Model (SAM) maps these concentrations onto interfacial properties at fluid-fluid (FF) and solid-fluid (SF) interfaces. At the FF interface we adopt a single-site Langmuir adsorption at constant temperature, which sets the fractional coverage, while a Gibbs-type relation links interfacial excess to the interfacial tension.

\subsubsection{Interfacial tension based on concentration near the fluid-fluid interface}

\begin{table*}
\centering
\includegraphics[width=1.0\linewidth]{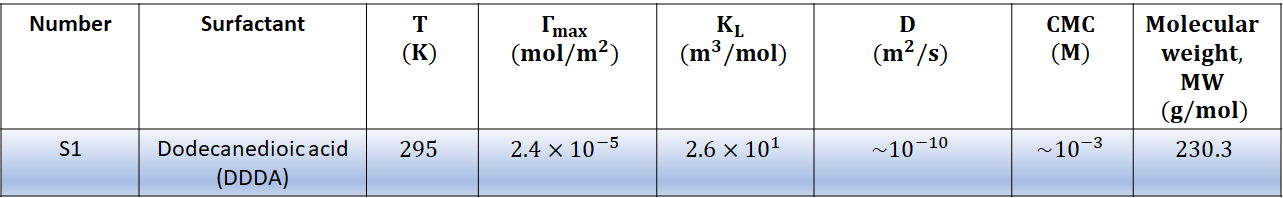}
\caption{Langmuir adsorption and transport parameters for the surfactant used in the simulations \cite{Chang1995}. For brevity, we refer to the surfactant by its number (S1).}
\label{table:2} 
\end{table*}

To capture IFT reduction at FF interfaces, we use the Langmuir isotherm (single-site occupancy, isothermal):
\begin{equation}
\mathrm{\frac{\theta}{1-\theta}=K_L C_{IFF}},
\label{eq:Langmuir}
\end{equation}
where $\theta$ is the fractional surface coverage and $\mathrm{K_L}$ is the Langmuir equilibrium constant. Relating $\sigma$ to $\mathrm{C_{IFF}}$ via a Gibbs-type expression gives
\begin{equation}
\mathrm{\frac{d\sigma}{dC_{IFF}}=-R T\,\Gamma=-R T\,\theta\,\Gamma_{max}},
\label{eq:Gibbs1}
\end{equation}
where $\mathrm{R}$ is the gas constant, $\mathrm{T}$ the temperature, $\Gamma$ the interfacial excess at coverage $\theta$, and $\Gamma_{max}$ the saturation excess. Substituting Eq.~\ref{eq:Langmuir} yields
\begin{equation}
\mathrm{\frac{d\sigma}{dC_{IFF}}=-R T\,\Gamma_{max}\!\left(\frac{K_L C_{IFF}}{1+K_L C_{IFF}}\right)}.
\label{eq:Gibbs2}
\end{equation}
Integration gives
\begin{equation}
\mathrm{\sigma_1-\sigma_0=-R T\,\Gamma_{max}\,\ln\!\left(\frac{1+K_L C_{IFF1}}{1+K_L C_{IFF0}}\right)}.
\label{eq:LG1}
\end{equation}
With $\mathrm{C_{IFF0}}=0$ at $\mathrm{t}=0$, Eq.~\ref{eq:LG1} reduces to
\begin{equation}
\mathrm{\sigma_1-\sigma_0=-R T\,\Gamma_{max}\,\ln\!\left(1+K_L C_{IFF1}\right)}.
\label{eq:LG2}
\end{equation}

\subsubsection{Contact angle based on concentration near the solid-fluid interface}

Adsorption at SF interfaces drives a wettability shift with respect to the invading phase, from non-wetting to wetting, and specifically for the simulated condition, from hydrophobic to hydrophilic. We employ a phenomenological relation:
\begin{equation}
\mathrm{\alpha_{ItSF}=\alpha_{I0SF}-\frac{C_{ISF}\,MW}{\rho}\times 100},
\label{eq:CA1}
\end{equation}
where $\alpha_{\mathrm{I0SF}}$ is the initial contact angle for the interface $\mathrm{I}$, $\alpha_{\mathrm{ItSF}}$ is the angle after time $\mathrm{t}$, $\mathrm{MW}$ is the surfactant molecular weight, and $\rho$ is the density of water. Throughout, the contact angle is measured through the invading (water) phase.

Finally, the updated interfacial tension and contact angle enter the capillary threshold through $\sigma\cos\alpha$ in Eq.~\ref{eq:Laplace}, which governs secondary-invasion decisions in the network.

\begin{table*}
\centering
\includegraphics[width=0.5\linewidth]{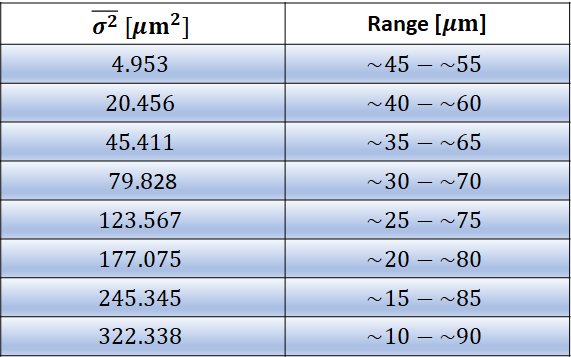}
\caption{Levels of pore-throat size heterogeneity used to generate the ensembles. Each row lists the prescribed variance $\overline{\mathrm{\sigma^2}}$ (in $\mu\mathrm{m}^2$) of the Gaussian throat distribution at fixed mean $\bar r \approx 50\,\mu\mathrm{m}$. The ``Range'' column reports the approximate span $\bar r \pm 1.96\sigma$ (i.e., the $\sim95\%$ interval), providing an intuitive measure of heterogeneity.}
\label{table:3} 
\end{table*}

\subsection{Coupling of PNM, STM, and SAM}

\begin{figure*}
\centering
\includegraphics[width=1.0\linewidth]{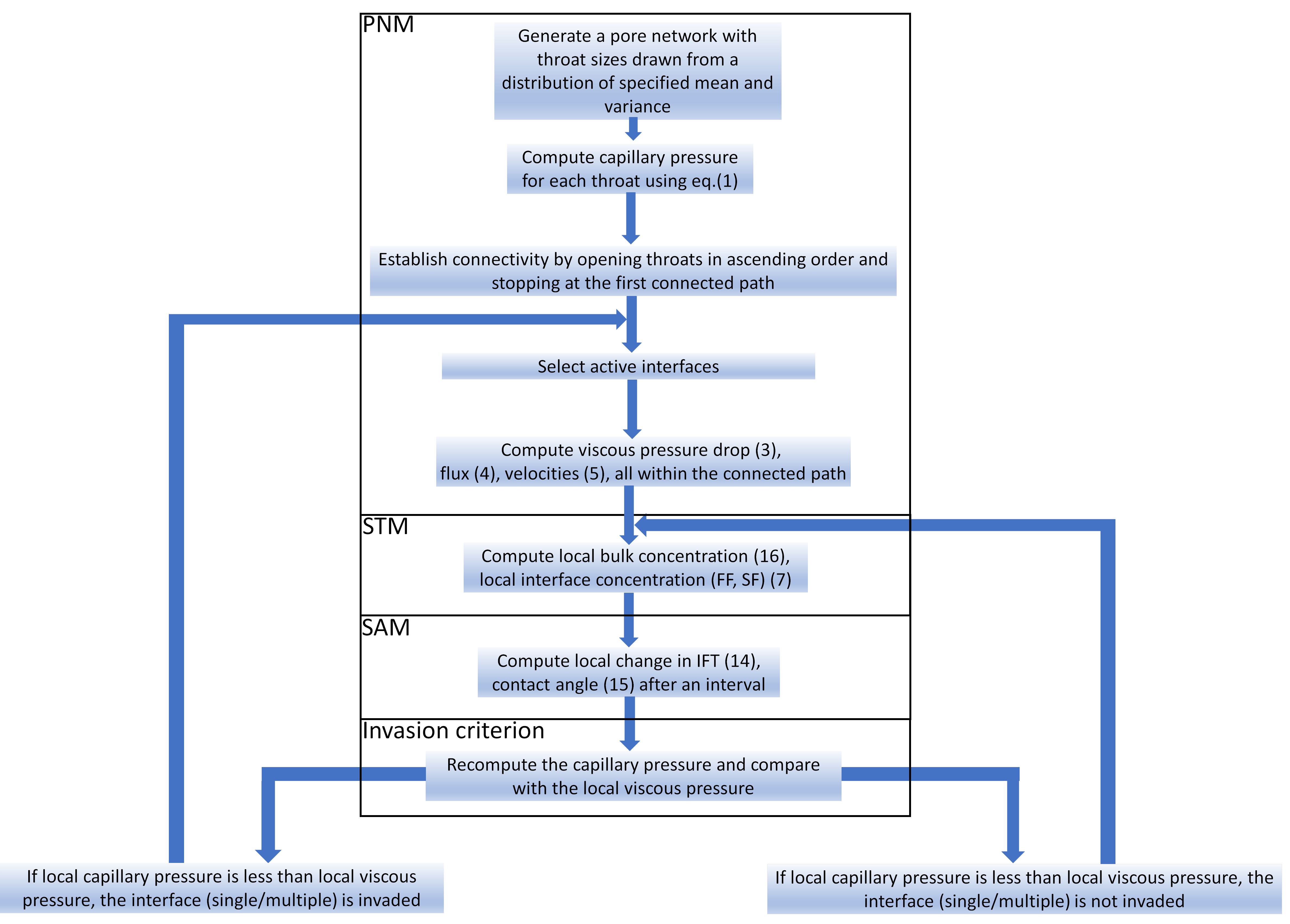}
\caption{Workflow linking primary invasion and hydrodynamics (PNM), surfactant transport (STM), and adsorption (SAM) to resolve surfactant-driven secondary invasion.}
\label{fig:2}
\end{figure*}

The three models are integrated as follows: The pore-network hydrodynamics provide edge fluxes $\mathrm{Q_{ij}}$ along the invaded path. These fluxes set the nodewise bulk concentration field $\mathrm{C_B}$ through a steady network mass balance. STM then yields the near-interface concentration $\mathrm{C_I}$ at each active throat via the film closure, while SAM converts the interface-specific concentrations to interfacial properties: the interfacial tension $\sigma$ through the Gibbs-Langmuir relation and the contact angle $\alpha$ through the wettability condition. Substituting $\sigma(t)$ and $\alpha(t)$ into the Laplace expression (Eq.~\ref{eq:Laplace}) updates the local capillary threshold $\mathrm{p_I(t)}$. An interface is advanced whenever the invading-phase nodal pressure meets or exceeds this threshold. The displaced (oil) phase is taken as hydrodynamically inactive, consistent with quasi-static IP.

We use the following network-level loop (also summarized in Fig.~\ref{fig:2}), referring to the ensuing dynamics as \emph{secondary invasion}:

\begin{enumerate}
  \item \textbf{Initialize concentrations.} At $\mathrm{t}=0$, set a uniform bulk concentration $\mathrm{C_{B0}}=2\,\mathrm{mM}$. On the currently invaded subnetwork, enforce nodewise advective mass balance
 \begin{equation}
  \sum_{\mathrm{j}}\!\left(\mathrm{C}_{\mathrm{B}\mathrm{j}}\,\mathrm{Q}_{\mathrm{i}\mathrm{j}}
  - \mathrm{C}_{\mathrm{B}\mathrm{i}}\,\mathrm{Q}_{\mathrm{i}\mathrm{j}}\right) = 0,
  \label{eq:CBalance}
 \end{equation}

  to obtain $\mathrm{C_{Bi}}$ at every node $\mathrm{i}$.
  
  \item \textbf{Update interfacial state.} From $\mathrm{C_B}$, we evaluate the near-interface concentration using the STM,
  \[
  \mathrm{C}_{\mathrm{I}} \;=\; \mathrm{C}_{\mathrm{B}}\!\left(1-\exp\!\left[-\frac{\mathrm{k}}{\mathrm{L}_{\mathrm{I}}}\,\mathrm{t}\right]\right)
  \]

  (Eq.~\ref{eq:MT2}). Use SAM to update interfacial properties: apply the Gibbs-Langmuir relation for $\sigma$ (Eqs.~\ref{eq:LG1}-\ref{eq:LG2}) and the SF closure for $\alpha$ (Eq.~\ref{eq:CA1}).
  
  \item \textbf{Invasion check and advance.} Form the updated capillary threshold $\mathrm{p_I(t)}$ by inserting $\sigma(\mathrm{t})$ and $\alpha(\mathrm{t})$ into Eq.~\ref{eq:Laplace}. For each active interface $\mathrm{I}$, advance it if
  \[
    \mathrm{P}_{\mathrm{I}}(\mathrm{t}) \;\ge\; \mathrm{p}_{\mathrm{I}}(\mathrm{t}),
  \]

  where $\mathrm{P_I(t)}$ is the invading-phase nodal pressure from the hydrodynamic pressure solver. If any interface advances, the algorithm updates the invaded set and returns to step~1. This step is iterated until no interfaces satisfy the criterion.
\end{enumerate}

This feedback between PNM, STM, and SAM continues until the system reaches a state where no further secondary-invasion events occur on the network considered.

\subsection{Analysis}
\label{analysis}

Following the primary displacement, surfactant activity gives rise to a secondary invasion phase, during which the invaded region continues to grow. At any time $t$, the invaded fraction is
\[
\mathrm{f(t)} \;=\; \frac{\mathrm{N}_{\mathrm{inv}}(\mathrm{t})}{\mathrm{N}_{\mathrm{T}}},
\qquad \mathrm{N}_{\mathrm{T}} = 1{,}012.
\]

where $\mathrm{N_{inv}(t)}$ is the number of invaded interfaces and $\mathrm{N_T}$ is the total number of pore throats in the network.

\paragraph{Invaded fraction fit per realization.}
For each realization $\mathrm{p}$, we correlate the time dependence of the invaded fraction with a shifted-and-scaled Gaussian cumulative distribution:
\begin{equation}
\mathrm{f_{pt}(t) \;=\; \big( f_{p\infty} - f_{p0} \big)\,
\Phi\!\left(\frac{t_{p} - \alpha_\mathrm{p} \overline{\tau_{p}}}{{\beta_\mathrm{p}\hat \tau_{p}}}\right) + f_{p0}}, 
\qquad \mathrm{p}=1,2,\ldots,300,
\label{eq:21}
\end{equation}
where $\mathrm{f_{p0}}$ is the invaded fraction at percolation (onset of secondary invasion), $\mathrm{f_{p\infty}}$ is the long-time value, $\Phi(\cdot)$ is the Gaussian CDF, $(\alpha_\mathrm{p},\beta_\mathrm{p})$ are optimization parameters with ensemble averaged values in the range [0.5,1.0] and $(\overline{\tau}_\mathrm{p},\hat{\tau}_\mathrm{p})$ are the realization’s interfacial mass-transfer statistics  defined below.

\paragraph{Interfacial mass-transfer timescales.}
For each interface, $\mathrm{I}$, we define the mass-transfer timescale as
\begin{equation}
\mathrm{\tau_I \;=\; \frac{\left(\frac{V}{S}\right)_I}{k_I}},
\label{eq:17}
\end{equation}
with $\mathrm{(V/S)_I}$ the local volume-to-area ratio (i.e., $\mathrm{L_{\mathrm{I}}}$) and $\mathrm{k_I}$ the local mass-transfer coefficient at the onset of secondary invasion.

The realization-level statistics are then
\begin{equation}
\mathrm{\overline{\tau}_p \;=\; \frac{1}{n}\, \sum_{I=1}^{n} \left(\frac{\left(\frac{V}{S}\right)_I}{k_I}\right)},
\qquad
\mathrm{\hat{\tau}_p \;=\; \sqrt{\frac{1}{n}\, \sum_{I=1}^{n} \left[\left(\frac{\left(\frac{V}{S}\right)_I}{k_I}\right) - \overline{\tau}_p\right]^2}},
\end{equation}
where $n$ is the number of interfaces included in the onset set for realization $p$. Here $\overline{\tau}_p$ and $\hat{\tau}_p$ denote the mean and standard deviation of $\{\tau_I\}$ for that realization.

\paragraph{Ensemble averages.}
Each ensemble contains a large number of realizations (200-300). The ensemble-mean timescale is
\begin{equation}
\overline{\overline{\tau}} \;=\; \frac{1}{n}\sum_{p=1}^{n}\overline{\tau}_{p}
\quad\text{where } n \in \{200,\ldots,300\}.
\end{equation}

\paragraph{Notation.}
Overbars on a single index (e.g., $\overline{\tau}_p$) indicate realization-level means; a double overbar (e.g., $\overline{\overline{\tau}}$) indicates an average across the ensemble. The hat on $\hat{\tau}_p$ denotes the realization-level standard deviation of $\{\tau_I\}$.

\section{Results and Discussion}

\begin{figure}[tbp]
\centering
\includegraphics[width=0.58\linewidth]{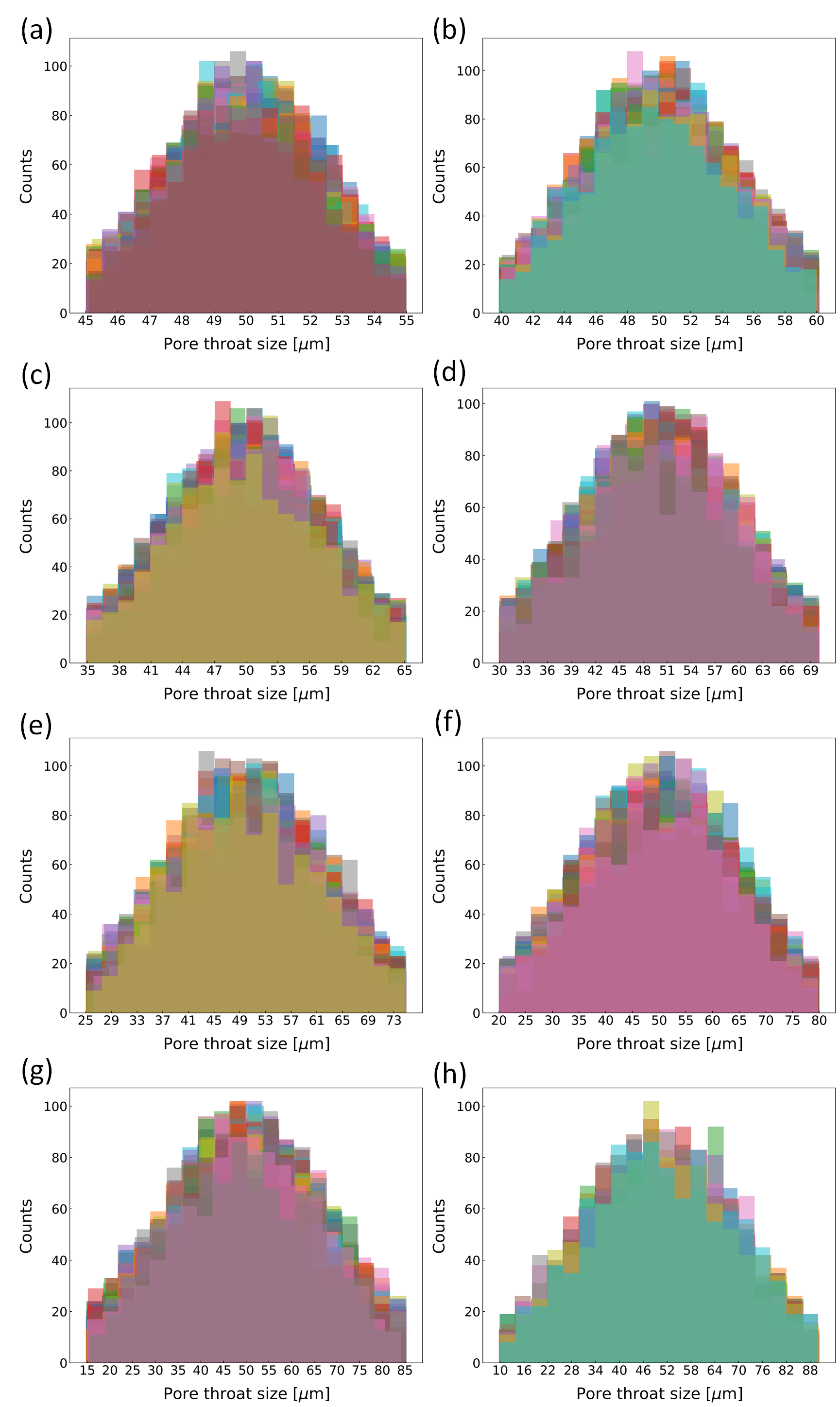}
\caption{Histograms of Gaussian pore-throat radii with similar means and different variances. Reported ensemble mean radius \(\overline{\overline{r}}\) and ensemble variance \(\overline{\sigma^2}\) are:
(a) \((49.998\,\mu\mathrm{m},\,4.953\,\mu\mathrm{m}^2)\),
(b) \((49.996\,\mu\mathrm{m},\,20.456\,\mu\mathrm{m}^2)\),
(c) \((49.986\,\mu\mathrm{m},\,45.411\,\mu\mathrm{m}^2)\),
(d) \((50.045\,\mu\mathrm{m},\,79.828\,\mu\mathrm{m}^2)\),
(e) \((50.049\,\mu\mathrm{m},\,123.567\,\mu\mathrm{m}^2)\),
(f) \((50.007\,\mu\mathrm{m},\,177.075\,\mu\mathrm{m}^2)\),
(g) \((49.986\,\mu\mathrm{m},\,245.345\,\mu\mathrm{m}^2)\),
(h) \((49.980\,\mu\mathrm{m},\,322.338\,\mu\mathrm{m}^2)\).}
\label{fig:6}
\end{figure}

We examine the spatio-temporal dynamics of fluid flow in a heterogeneous pore network. The extent of heterogeneity is quantified through the pore throat size distribution, defined by a Gaussian distribution with a mean radius ($\overline{r}$) and prescribed variance. A higher variance signifies stronger heterogeneity within the network structure. To represent this variability, we construct 250-300 independent realizations sampled from the same underlying distribution. While individual networks display modest differences in their average radius and spread, the collective ensemble reflects the prescribed mean and variance of the Gaussian distribution. A statistical summary is presented in Table~\ref{table:3}, and representative network configurations are displayed in Fig.~\ref{fig:6} in Section~\ref{section:Supplementary}.

\begin{figure*}[!b]
\includegraphics[width=0.8\linewidth]{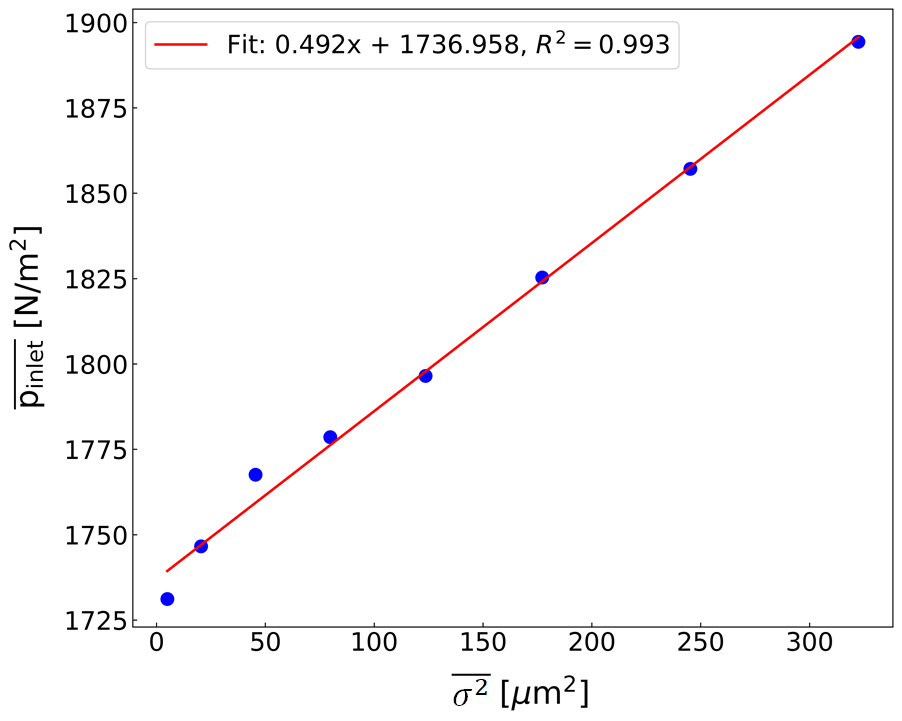}
\caption{Mean inlet pressure, $\mathrm{p_{inlet}}$ plotted with mean variance, $\overline{\mathrm{\sigma^2}}$ for $\mathrm{C_B}=2$ $\mathrm{mM}$ and $\mathrm{S1}$ parameters.} 
\label{fig:8} 
\end{figure*}

Secondary drainage is analyzed following a primary drainage event in which clean water (no surfactant) displaces oil in an hydrophobic, oil-saturated network. 
After breakthrough, we switch to surfactant-bearing water at the same inlet pressure to probe adsorption-driven wettability shift and continued invasion. The presence of surfactants lowers the interfacial tension and changes wettability at the oil-water boundary, thereby promoting displacement driven by pressure differences across pore throats. In these simulations, we define the network entry pressure as the smallest capillary pressure at which a continuous pathway from inlet to outlet exists. Once this baseline invasion pattern is established, the invading phase is switched to surfactant-laden water, and subsequent displacement is tracked to capture the additional advance caused by interfacial tension reduction and wettability change. Because not all pathways are activated by the surfactant, the invasion sequence continues until two conditions are satisfied: (i) a connected pathway exists between inlet and outlet and (ii) further advance is observed due to the surfactant effect. 
At this point, the inlet pressure is fixed and the associated invaded fraction is designated as the percolation threshold value. For clarity, we refer to the initial advance in the absence of surfactant as the \textit{primary invasion} and to the subsequent displacement triggered by surfactant adsorption as the \textit{secondary invasion}. 
Using this procedure, we find that the required inlet pressure exhibits a linear increase with the degree of heterogeneity, as summarized in Fig.~\ref{fig:8} in Section~\ref{section:Supplementary}.

\begin{figure*}
\centering
\includegraphics[width=0.8\linewidth]{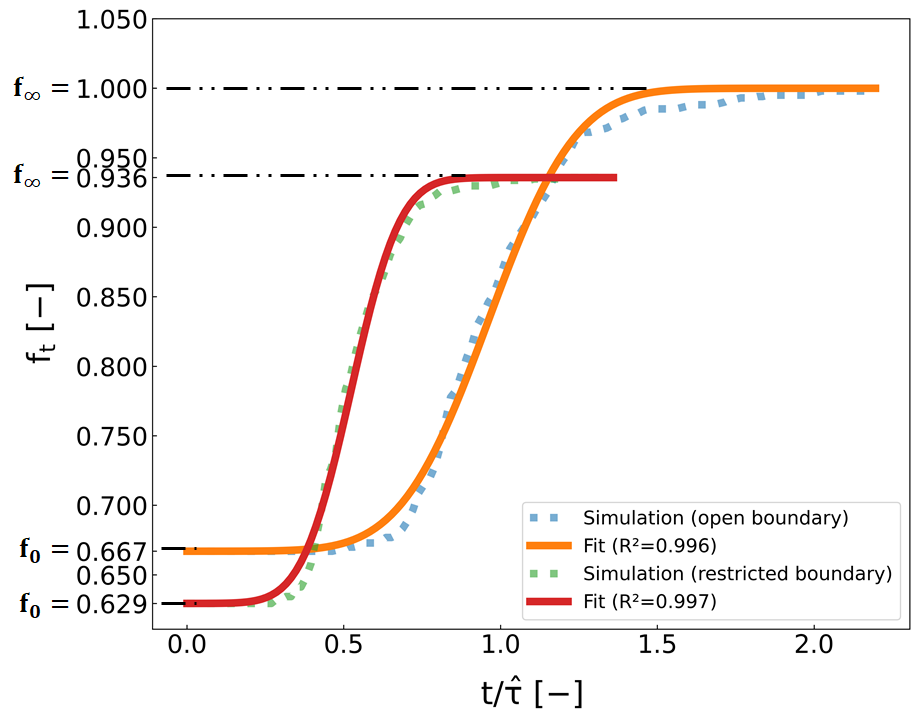}
\caption{Representative invaded fraction trajectories under open and restricted boundaries 
conditions, $\mathrm{f_t}$, together with their corresponding Gaussian CDF (error-function) fits as a function 
of time scaled with the median mass-transfer timescale. Results are obtained from an ensemble of 279 realizations generated from a PDF 
with a prescribed mean pore throat size of $50.0\,[\mu\text{m}]$ and variance of $4.94\,[\mu\text{m}^2]$. 
The specific network illustrated here has a mean pore throat size of $49.95\,[\mu\text{m}]$ and 
a variance of $5.14\,[\mu\text{m}^2]$. 
The temporal behavior reflects progressive surfactant adsorption 
at the oil-water interface (bulk concentration $\mathrm{C_B}=2\,\text{mM}$, surfactant S1), which drives successive invasion events. The invasion reaches different long-time limits under the two boundary conditions: complete displacement for the open case ($\mathrm{f_{\infty}}=1.0$) and partial 
saturation for the restricted case ($\mathrm{f_{\infty}}=0.936$). The representative fits capture the dynamics with high accuracy ($\mathrm{R^2}=0.996,\,0.997$), as further discussed in Section~\ref{analysis}. Since the invasion dynamics are qualitatively similar, subsequent analysis 
in this paper focuses on the open boundary condition.}
\label{fig:3} 
\end{figure*}

Fig.~\ref{fig:3} presents the time-dependent progression of invasion, expressed through the invaded fraction. The simulation begins from the initial percolation threshold, obtained by prescribing a starting invasion pattern, and uses the surfactant parameter set $\mathrm{S1}$ (Table~\ref{table:1}). 
The onset of secondary invasion occurs at \(\mathrm{t}/\mathrm{\hat \tau} \approx 0.5\), marked by the invaded fraction surpassing the percolation threshold \(\mathrm{f}_0 = 0.667\). Beyond this point, surfactants continue to accumulate both at pre-existing fluid-fluid interfaces and at new interfaces created as the displacement advances. Each time a pore throat is invaded through this mechanism-driven by the reduction of interfacial tension relative to the local capillary pressure and change in wettability- the mass transfer coefficient $k$ is recalculated for the remaining active interfaces.  

The two boundary conditions considered-open and restricted-yield different steady-state invaded fractions, as shown in the figure. 
Under open boundary conditions, the invasion proceeds until full displacement is achieved ($\mathrm{f_{\infty}} = 1.0$), whereas the restricted case saturates at a lower value ($\mathrm{f_{\infty}} = 0.936$) due to confinement effects. The respective Gaussian CDF (error-function) fits 
reproduce the simulation data with excellent agreement, yielding coefficients of determination of $\mathrm{R^2} = 0.996$ (open boundary) and $\mathrm{R^2} = 0.997$ (restricted boundary). 
These results highlight the role of boundary conditions in controlling the ultimate extent of invasion, even when the early-time growth dynamics are similar.  
Since both cases follow the same qualitative trend, in the remainder of this paper, we focus on the open boundary condition for clarity and consistency.

In our simulations, the pore throat sizes are assigned according to a Gaussian distribution with specified mean and variance. This distribution governs both the local capillary entry pressures and the adsorption behavior of surfactants, since the velocity of the invading phase is determined through the Hagen-Poiseuille relation (Eq.~\ref{eq:Flux}). Consequently, the gradual increase in the invaded fraction can be understood as a continuous update of the underlying 
velocity field. Under the imposed boundary condition of a fixed pressure drop, each additional invasion lowers the effective resistance of the network and enhances the total flux.  

To describe the spatio-temporal growth of the invasion within such Gaussian-distributed networks, we employ the error function-the cumulative distribution function (CDF) of the Gaussian probability density function-as a predictive model for the displacement dynamics of each realization (Eq.~\ref{eq:21}). 
In addition, we define a characteristic mass transfer timescale that becomes relevant in the secondary invasion stage, expressed as the ratio of a 
characteristic length to the mass transfer coefficient (Eq.~\ref{eq:17}). 
The length scale is represented by the inverse of the surface-area-to-volume ratio, which provides a geometrical measure linked to network permeability and serves as a proxy for flux capacity. 
Dividing this length by the mass transfer coefficient yields a quantitative estimate of the rate at which surfactant transfers from the bulk solution to the interfaces. 
When this timescale is incorporated into the Gaussian CDF formulation, it provides a compact description of how the invaded fraction evolves in time under a known set of parameters. This sigmoidal (error-function-like) form is consistent with classical cumulative pore-size arguments in porous media where a lognormal distribution of throat sizes naturally produces an error-function-type saturation curve \cite{Kosugi1994,Kosugi1996}. Furthermore, in our previous work, we showed that the same collapse can be obtained in two equivalent parameterizations \cite{Bhattacharjee2025}: (i) time, scaled by an interfacial mass-transfer timescale that reflects surfactant transport and adsorption, and (ii) the progressive relaxation of the local Laplace-pressure entry threshold at fixed inlet pressure. In both approaches, the invaded fraction follows the same sigmoidal trajectory, indicating that mass-transfer-driven interfacial conditioning and capillary-threshold relaxation are two views of the same secondary invasion process.

Because individual realizations exhibit distinct percolation thresholds while converging to the same steady-state invaded fraction, all parameters are evaluated at the end of the primary invasion 
stage to properly incorporate the initial condition. As shown in Fig.~\ref{fig:3}, the fitting model (Eq.~\ref{eq:21}) captures the invasion dynamics well for the representative realization displayed. Since each network is generated independently, even though they share the same parent distribution, the error function integral cannot be expressed in closed form and must instead be computed numerically for each realization.  

\begin{figure*}[tbp]
    \centering
    \includegraphics[width=0.6\linewidth]{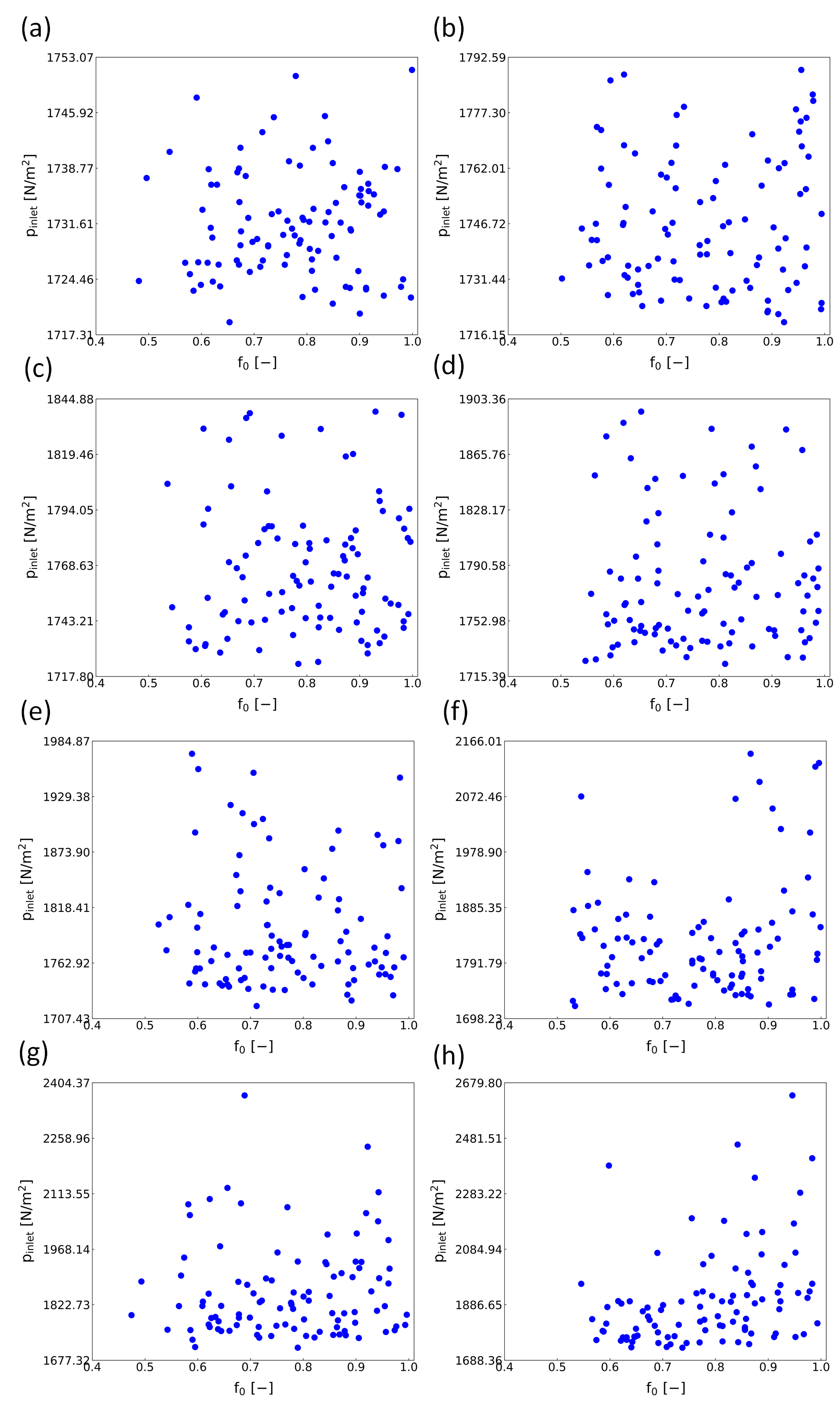}
    \caption{Percolation threshold \(f_0\) versus inlet pressure for ensembles with similar mean pore-throat radius and different variances. Reported ensemble mean \(\overline{\overline{r}}\) and variance \(\overline{\sigma^2}\) are:
        (a) \((49.998\,\mu\mathrm{m},\,4.953\,\mu\mathrm{m}^2)\),
        (b) \((49.996\,\mu\mathrm{m},\,20.456\,\mu\mathrm{m}^2)\),
        (c) \((49.986\,\mu\mathrm{m},\,45.411\,\mu\mathrm{m}^2)\),
        (d) \((50.045\,\mu\mathrm{m},\,79.828\,\mu\mathrm{m}^2)\),
        (e) \((50.049\,\mu\mathrm{m},\,123.567\,\mu\mathrm{m}^2)\),
        (f) \((50.007\,\mu\mathrm{m},\,177.075\,\mu\mathrm{m}^2)\),
        (g) \((49.986\,\mu\mathrm{m},\,245.345\,\mu\mathrm{m}^2)\),
        (h) \((49.980\,\mu\mathrm{m},\,322.338\,\mu\mathrm{m}^2)\).
        Parameters for surfactant S1 are used.}
    \label{fig:9}
\end{figure*}

To demonstrate the predictive capacity of this approach, we analyze ensembles of networks with varying degrees of heterogeneity and compare their corresponding percolation thresholds (Fig.~\ref{fig:9} in Section~\ref{section:Supplementary}). Because each network is generated stochastically, the realized mean and variance of its structural parameters differ slightly from the prescribed values which is illustrated in Fig.~\ref{fig:6} in Section~\ref{section:Supplementary}.

For each network realization, we apply the same analysis of surfactant-driven displacement using the parameter set S1 at the specified bulk concentration. The percolation threshold varies from one realization to the next, reflecting the stochastic placement of pore throat sizes. 
Despite these differences at onset, all cases approach a terminal invaded fraction of unity at steady states, indicating that the initial condition does not set the ultimate extent of invasion. 
By design, the algorithm yields $\mathrm{f(t)\to 1}$. 
Importantly, relaxing this constraint and permitting terminal values below unity does not change the observation that the invasion trajectory is well captured by the Gaussian CDF (error-function) form. 
In addition, we find no observable statistical relationship between the required inlet driving pressure and the percolation threshold (Fig.~\ref{fig:8} in Section~\ref{section:Supplementary}).

\begin{figure*}
    \centering
    \includegraphics[width=0.85\linewidth]{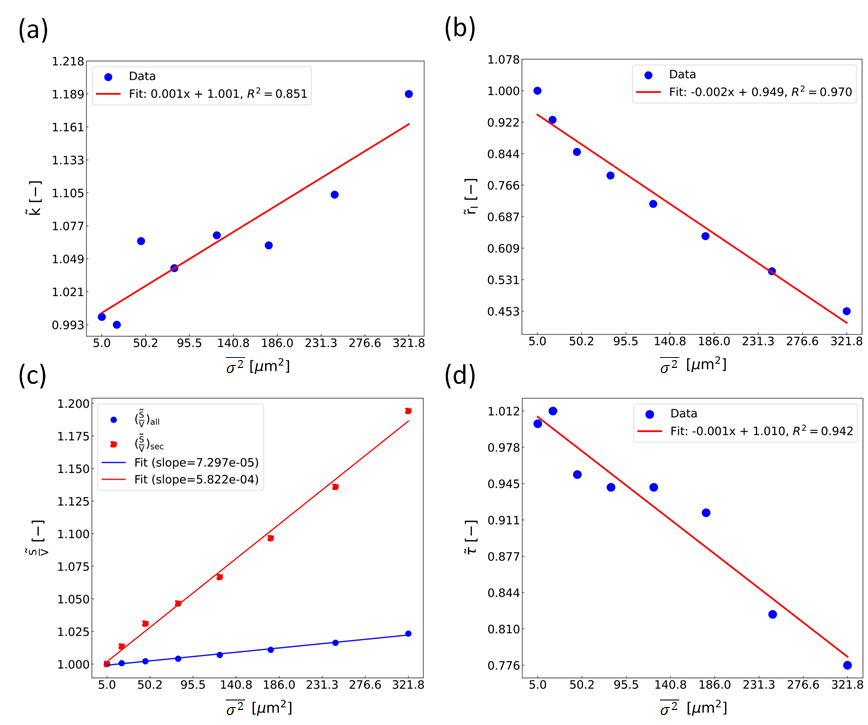}
    \caption{ (a) Normalized mass transfer coefficient for the ensemble, \(\mathrm{\tilde{k}}\),
    (b) normalized interface size of the ensemble, \(\mathrm{\tilde{r}_I}\),
    (c) normalized surface-area-to-volume ratio of the ensemble, \(\widetilde{\dfrac{\mathrm{S}}{\mathrm{V}}}\), and
    (d) normalized mass transfer timescale, \({\tilde{\tau}}\),
    calculated for the S1 surfactant-driven invasion dynamics with different prescribed ensemble variance, \(\overline{\mathrm{\sigma^2}}\), as indicated by the x-axis.
    The subscripts `all' and `sec' in the legend denote the ratio for all pore throat sizes and interface sizes at the onset of secondary invasion, respectively. At fixed inlet pressure, the surfactant lowers effective capillary thresholds (\(\sigma\cos\alpha\)), so larger \(\overline{\mathrm{\sigma^2}}\) promotes earlier activation of high-conductance throats and faster connectivity of the invading pathway. As a result, \(\mathrm{\tilde{k}}\) increases with \(\overline{\mathrm{\sigma^2}}\), while the interface sizes decreases, yielding a higher \(\widetilde{\dfrac{\mathrm{S}}{\mathrm{V}}}\)-especially at the onset of secondary invasion-which enhances interfacial exchange. The combined effect is a systematic reduction in the mass transfer timescale.   \textbf{Normalization definition:} for any quantity $q$ plotted in panels (a-d),
    $\tilde{q} \equiv \overline{q} / \overline{q}\big|_{\overline{\mathrm{\sigma^2}}=\overline{\mathrm{\sigma^2}}_{\min}}$,
    i.e., each value is divided by its value in the baseline case with the
    \emph{smallest prescribed ensemble variance} $\overline{\mathrm{\sigma^2}}_{\min}$
    (here, $4.953~\mu\mathrm{m}^2$).}
    \label{fig:4} 
\end{figure*}

\begin{figure*}
\centering
\includegraphics[width=0.8\linewidth]{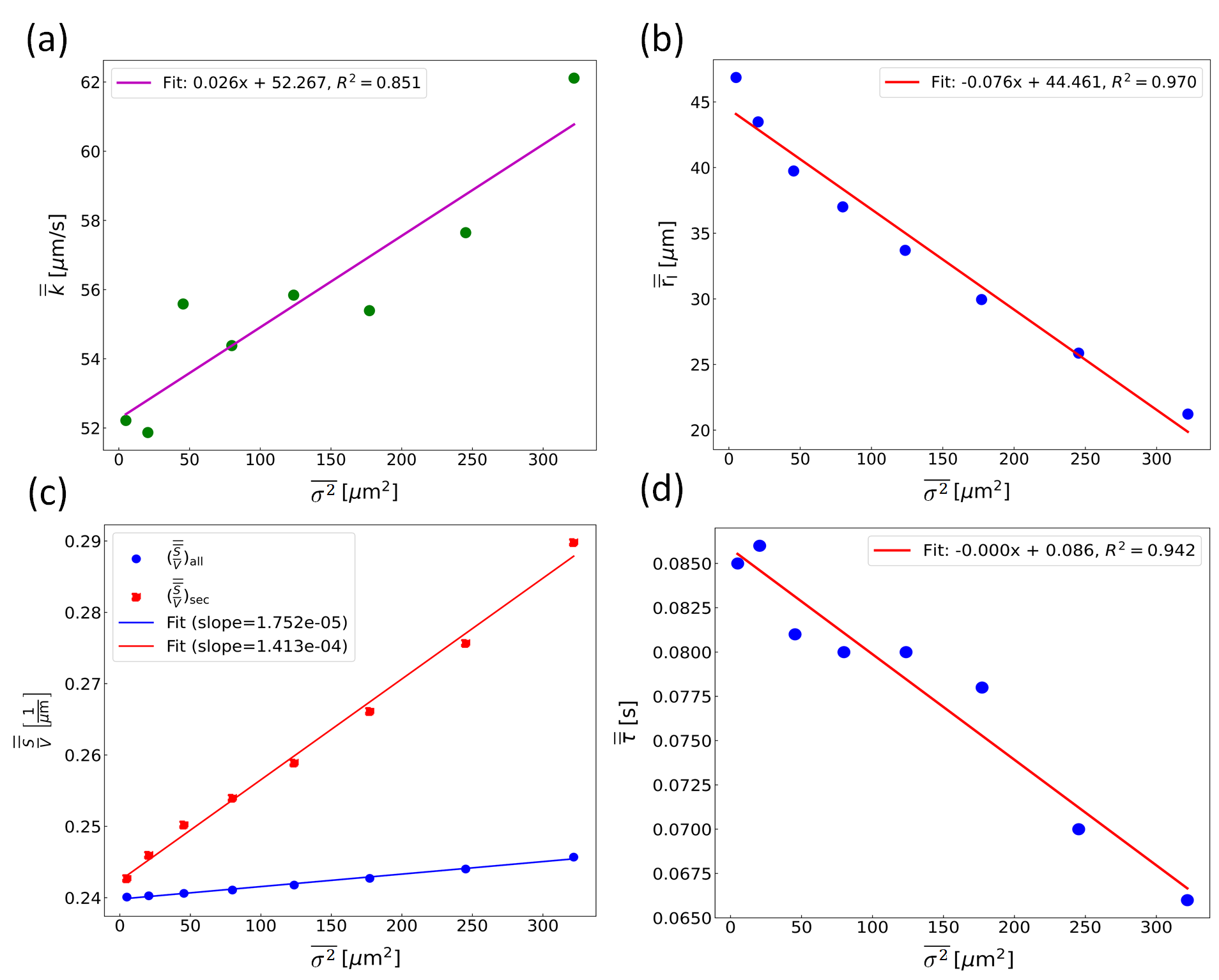}
\caption{(a) Mass transfer coefficient of the ensemble, \(\mathrm{\overline{\overline{k}}}\), (b) interface size of the ensemble, \(\mathrm{\overline{\overline{r_I}}}\), (c) Surface area to volume ratio of the ensemble, \(\mathrm{\overline{\overline{\frac{S}{V}}}}\), and (d) mass transfer timescale, ${\overline{\overline{\tau}}}$ are plotted with the variance of the ensemble, $\overline{\sigma^2}$. The subscripts `all' and `sec' in the legend denote the ratio for all pore throat sizes and interface sizes at the onset of secondary invasion, respectively, in the case of open boundaries only. Parameters of surfactant S1 are considered.}
\label{fig:10} 
\end{figure*}

The key parameters used to fit the surfactant-driven invasion can be regarded as characteristic parameters for the ensemble and serve as scaling factors between the various heterogeneity levels, marked by the variances. 
Following this comparison, we observe that the normalized mass transfer coefficient of the ensemble increases linearly as we increase the variance of the pore distribution, though at a much slower rate compared to the interface size, which is evident from the slopes (Fig.~\ref{fig:4}(a),(b)). 
This can be attributed to the fact that for higher variances, the water traverses through smaller interface sizes and a larger surface area-to-volume ratio (Fig.~\ref{fig:4}(c)), which enhances the observed increase in mass transfer. 
Plotting the normalized mass transfer timescale, which is the inverse of the mass transfer, as a function of variance shows the expected linearly decreasing trend with the variance increase (Fig.~\ref{fig:4}(d)). 
This linear decrease suggests that as the heterogeneity of the system increases, the mass transfer rate improves, leading to shorter timescales. The reduction in timescale indicates that mass transfer occurs more rapidly in systems with higher variance, which can be attributed to the increased surface area-to-volume ratio for the initial invasion pattern. 
For the analysis in absolute values, the reader is referred to Fig.~\ref{fig:10} in Section~\ref{section:Supplementary}.

\begin{figure*}
    \centering
    \includegraphics[width=1.0\linewidth]{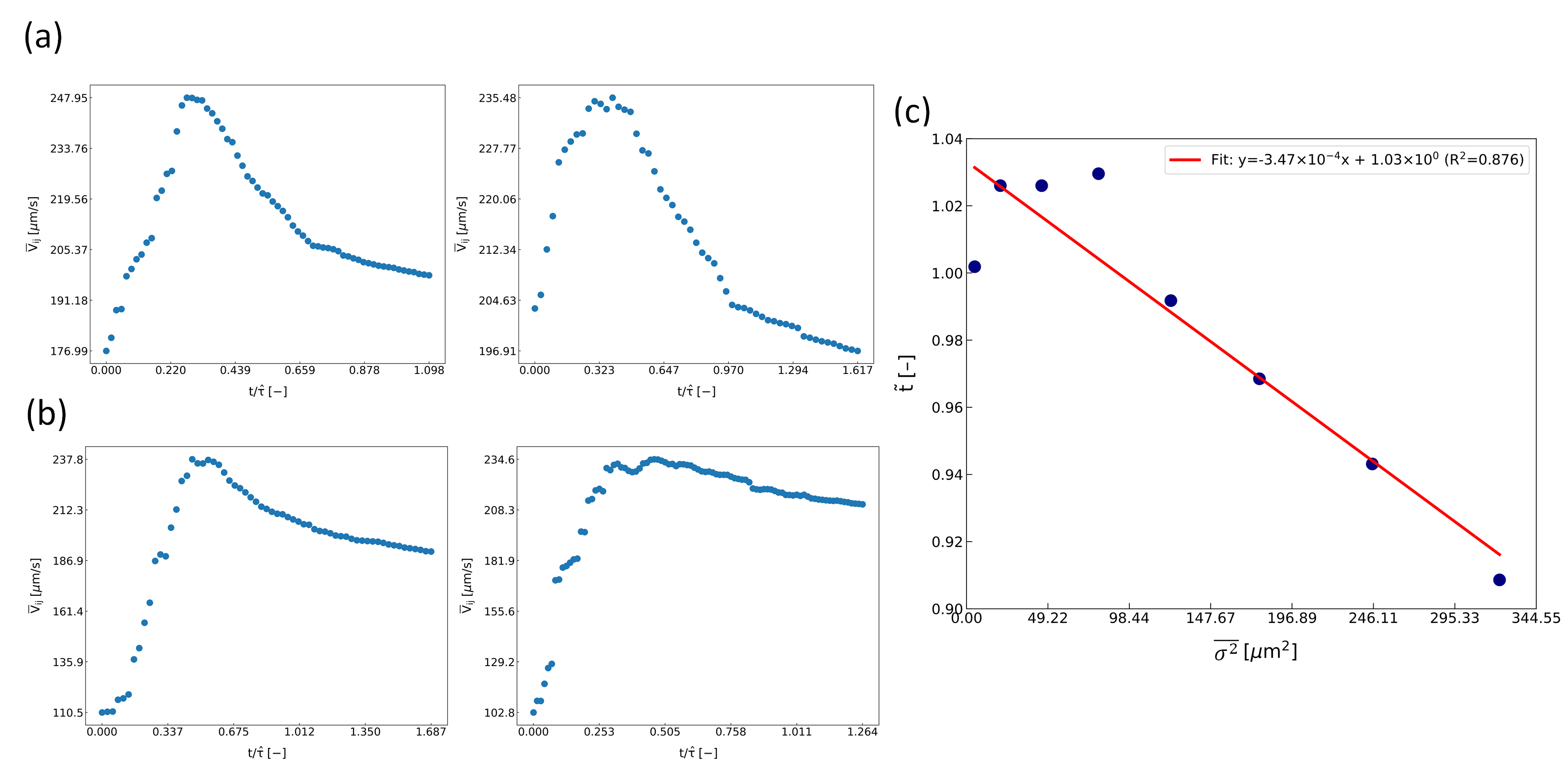}
    \caption{%
    {(a-b)} Mean velocities of the invaded paths, \(\mathrm{\overline{V}_{ij}}\) as a function of time scaled with the respective median mass-transfer timescale during S1 surfactant-driven invasion. Each panel shows two independent realizations from ensembles with mean variances, \(\overline{\mathrm{\sigma^2}}=\{4.953,\,322.338\}\,\mu\mathrm{m}^2\).
    The values rise to a peak and then decay as capillary entry events rapidly accelerate newly opened pathways and, once the backbone forms, viscous redistribution and geometric bottlenecks progressively reduce the local driving. The surfactant lowers the effective capillary thresholds (via \(\sigma\cos\alpha\)), sustaining invasion beyond the first breakthrough and imprinting the transient peak in \(\mathrm{\overline{V}_{ij}}\), 
    (c) Normalized full-sweep time, \(\mathrm{\tilde{t}}\), versus \(\overline{\mathrm{\sigma^2}}\); the red line is a linear fit with the displayed equation, showing that larger variance shortens the full-sweep time.}
    \label{fig:5} 
\end{figure*}

To further investigate the impact of heterogeneity, we examine the time evolution of the mean invaded-path velocities up to full sweep, defined as the time at which the defending oil phase is completely displaced by the invading water phase. 
Figs.\ref{fig:5}(a-b) show two independent realizations drawn from ensembles with prescribed variances \(\overline{\mathrm{\sigma^2}}=\{4.953,\,322.338\}\,\mu\mathrm{m}^2\), respectively. 
In both cases, the mean velocities exhibit a sharp rise followed by a gradual decay, reflecting the two dominant stages of displacement. 
The initial acceleration corresponds to the rapid mobilization of oil as the invading water-surfactant phase breaches low-threshold throats and accelerates through newly opened pathways. 
The subsequent decay arises as the flow consolidates into a stabilized backbone, reducing local pressure gradients and slowing the rate of invasion. 
To compare across heterogeneities, we examine the normalized full-sweep time, \(\tilde{\mathrm{t}}\), as a function of \(\overline{\mathrm{\sigma^2}}\). As shown in Fig.~\ref{fig:5}(c), \(\tilde{\mathrm{t}}\) decreases monotonically with increasing variance, indicating that broader pore-size distributions accelerate the invasion process. 
Physically, high-variance ensembles contain a wider spectrum of throat sizes, which increases the likelihood of early activation of large-conductance pathways. 
These preferential channels rapidly connect the inlet to the outlet, reduce the hydraulic resistance of the evolving backbone, and thereby shorten the overall duration required for the invading water-surfactant phase to fully displace the defending oil. 
In contrast, low-variance systems lack such early shortcuts, leading to more gradual pathway development and longer full-sweep times.

\section{Conclusion}

This study employs a reduced-order framework for surfactant-driven invasion in heterogeneous pore networks, integrating pore-network, transport, and adsorption modules. Using literature-based parameters for the S1 surfactant, we examine how adsorption-mediated changes in interfacial tension and wettability influence invasion dynamics.

To quantify the dynamics, we introduced a mass-transfer timescale, defined by an effective, lumped convection-diffusion transfer coefficient (units of velocity) and a heterogeneity-dependent geometric length scale. In networks with Gaussian-distributed pore radii, the temporal growth of the invaded fraction collapses onto the Gaussian CDF (error function), providing a compact descriptor of the invasion process.

A consistent physical picture emerges: At constant inlet pressure, chemical alterations driven by adsorption advance the invasion front, while structural heterogeneity dictates the rate. 
Increasing the variance promotes earlier activation of large-conductance throats, yielding a monotonic reduction in normalized full-sweep time. 
By redistributing activity toward smaller throats and opening high-flux conduits sooner, broader pore-size distributions increase the surface-area-to-volume ratio and expose more reactive interface-conditions that intensify interfacial exchange and accelerate secondary invasion.

Our results demonstrate that heterogeneity is not just geometric variation in the pore network; it directly regulates the rate of secondary invasion. Increasing the variance in throat sizes creates more small throats and increases the surface-area-to-volume ratio along the active flow pathway. This exposes more reactive interface and routes flow through narrower regions, which enhances interfacial mass transfer and shortens the characteristic mass-transfer timescale. As a result, heterogeneity acts as a control parameter: it selects the pathways available for surfactant transport and adsorption, and it sets the timescale over which the interface conditions evolve and the invading phase advances under a fixed inlet pressure. 
Since this framework currently neglects ganglion deformation, multi-ganglion interactions, and explicit trapping of isolated cluster, future extensions that incorporate these processes, along with richer adsorption kinetics, will be critical for predicting the onset of secondary invasion, the asymptotic invaded fraction, and the morphology of residual ganglia.

In summary, this work establishes a physically grounded, mass-transfer-based lens for analyzing surfactant-driven invasion in heterogeneous networks. 
By showing how structural heterogeneity modulates invasion dynamics relative to a common base case, it provides a scalable foundation for interpreting experiments, guiding design of enhanced recovery and remediation strategies, and extending pore-network modeling toward more realistic multi-physics settings.

\clearpage
\bibliography{Ref}

\clearpage
\section{Supplementary}
\label{section:Supplementary}
\clearpage
\subsection{Pore throat distribution}
\suppressfloats[t]

\clearpage
\subsection{Representative fits of all variances}
\suppressfloats[t]
\begin{figure}[tbp]
\centering
\includegraphics[width=0.64\linewidth]{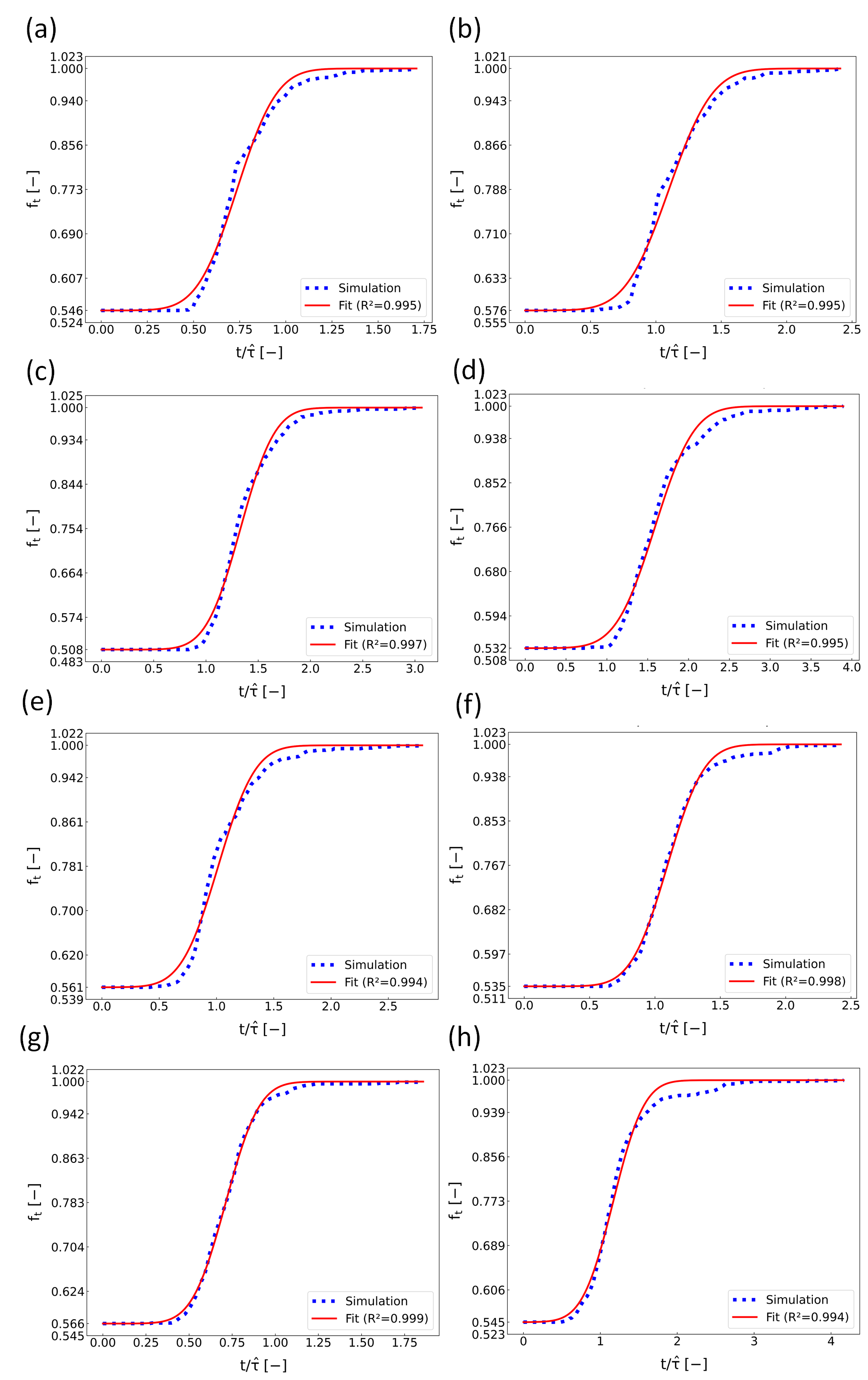}
\caption{Representative fit (highest \(R^2\) value) of the ensemble for different heterogeneities. Reported mean radius \(\overline{r}\) and variance \(\sigma^2\) pairs are:
(a) \((50.06\,\mu\mathrm{m},\,5.08\,\mu\mathrm{m}^2)\),
(b) \((49.08\,\mu\mathrm{m},\,21.45\,\mu\mathrm{m}^2)\),
(c) \((49.36\,\mu\mathrm{m},\,48.43\,\mu\mathrm{m}^2)\),
(d) \((49.67\,\mu\mathrm{m},\,78.95\,\mu\mathrm{m}^2)\),
(e) \((50.32\,\mu\mathrm{m},\,121.43\,\mu\mathrm{m}^2)\),
(f) \((49.35\,\mu\mathrm{m},\,176.10\,\mu\mathrm{m}^2)\),
(g) \((49.54\,\mu\mathrm{m},\,247.34\,\mu\mathrm{m}^2)\),
(h) \((50.43\,\mu\mathrm{m},\,316.33\,\mu\mathrm{m}^2)\).}
\label{fig:7}
\end{figure}

\clearpage
\subsection{Effect of heterogeneity on inlet pressure}
\suppressfloats[t]

\clearpage
\end{document}